\documentclass{article}
\usepackage{icrctc07}

\title{Ultra High Energy Cosmic Rays: origin and propagation}
\shorttitle{UHECR: Origin and propagation}

\authors{Todor Stanev} 
\shortauthors{Todor Stanev}
\afiliations{Bartol Research Institute, Department of Physics and
 Astronomy, University of Delaware, Newark, DE 19716, U.S.A.}  
\email{stanev@bartol.udel.edu}

\abstract{ We discuss the basic difficulties in understanding the origin
 of the highest energy particles in the Universe - the ultrahigh energy
 cosmic rays (UHECR). It is difficult to imagine the sources they are 
 accelerated in. Because of the strong attenuation of UHECR
 on their propagation from the sources to us these sources should be 
 at cosmologically short distance from us but are currently not
 identified. We also give information of the most recent experimental
 results including the ones reported at this conference 
 and compare them to models of the UHECR origin.
} 

\begin{document}
\maketitle

\section{Introduction}
  More than forty years ago, in 1963, John Linsley~\cite{Linsley63} 
 published an article about the detection of a cosmic ray of energy 
 10$^{20}$ eV. The article did not go unnoticed, neither it provoked
 many comments. The few physicists that were interested in high energy
 cosmic rays were at that time convinced that the cosmic ray energy spectrum
 will continue forever. The fact that cosmic rays may have energies
 exceeding 10$^6$ GeV (10$^{15}$ eV) was established in the late 
 thirties by Pierre Auger and his collaborators. Showers of higher
 and higher energy were detected in the mean time - seeing a
 10$^{20}$ eV shower seem to be only a matter of time and exposure. 
 Already in the fifties there was a discussion about the origin
 of such ultra high energy cosmic rays and Cocconi~\cite{Cocconi56}
 reached the conclusion that they must be of extragalactic origin
 since the galactic magnetic fields are not strong enough to contain such
 particles.

  How exclusive this event is became obvious three years later, after
 the discovery of the microwave background radiation (MBR). 
 Almost simultaneously Greisen in US~\cite{G} and
 Zatsepin\&Kuzmin~\cite{ZK} in the USSR published papers 
 discussing the propagation of ultra high energy particles in 
 extragalactic space. They calculated the energy loss distance
 of nucleons interacting in the microwave background and reached
 the conclusion that it is shorter than the distances between
 powerful galaxies. The cosmic ray spectrum should thus have an
 end around energy of 5$\times$10$^{19}$ eV. This effect is now
 known as the GZK cutoff. 

  The experimental statistics of such events grew with the years,
 although not very fast. The flux of UHECR of energy above 10$^{20}$ eV
 was estimated to be 0.5 to 1 event per square kilometer per century per
 steradian. Even big detectors of area tens of km$^2$ would only
 detect  few events for ten years of work. The topic became one of
 common interest during the last decade of the last century when
 ideas appeared for construction of detectors with effective areas in
 thousands of km$^2$~\cite{Auger,TA}. Such detectors would detect tens
 of events per year and finally solve all mysteries surrounding UHECR.
 The Auger observatory (3,000 km$^2$) is now almost fully deployed
 and the Telescope Array (TA, 1,000 km$^2$)  is being constructed in Utah,
 U.S.A. The expectations for the flux of UHECR
 are now smaller. The High Resolution Fly's Eye (HiRes) and Auger 
 have shown that the rate of events above 10$^{20}$ eV is about 
 10 times smaller than previously thought.

 Cosmic rays are defined as charged nuclei that originate
 outside the solar system. They come on a featureless, power law like,
 $F(E) = K \times E^{-\alpha}$, spectrum that extends beyond
 10$^{11}$ GeV per particle.
 There are only two distinct features in the whole spectrum.
 At energy above 10$^6$ GeV the power law index $\alpha$ steepens
 from 2.7 to about 3.1. This is called {\em the knee} of the cosmic
 ray spectrum. At energy above 3$\times$10$^9$ GeV the spectrum flattens
 again at {\em the ankle}.

 The common wisdom is that cosmic rays below the knee are accelerated 
 at galactic supernovae remnants. Cosmic rays above the knee are also
 thought to be of galactic origin, although there is no clue of their
 acceleration sites. Cosmic rays above the ankle are thought to be
 extragalactic.

 More recently, with the improved accuracy and exposure of the 
 modern experiments, several theoretical models that fit the 
 measured ultra high energy cosmic rays (UHECR) spectrum have 
 been developed. We will discuss the experimental data and compare
 them to some of these models.

\subsection{Air Shower Detection Methods.}

 Cosmic rays of energy above 10$^{14}$ eV are detected by the showers
 they generate in the atmosphere. The atmosphere contains more than
 ten interaction lengths even in vertical direction and is much deeper
 for particles that enter it under higher zenith angles. It is thus
 a deep calorimeter in which the showers develop, reach their maximum,
 and then start being absorbed. There are generally two types of air
 shower detectors: air shower arrays and optical detectors.
 Air shower arrays consist of numerous particle detectors
 that cover large area. The shower triggers the array by coincidental
 hits in many detectors. The most numerous particles in an air shower
 are electrons, positrons and photons. The shower also contains 
 muons, that are about 10\% of all shower particles, and hadrons.
 
 The direction of the primary particle can be reconstructed quite well
 from the timing of the different hits, but the shower energy requires
 extensive Monte Carlo work with hadronic interaction models that are
 extended orders of magnitude above the accelerator energy range.
 The main composition sensitive variable is the ratio of the number of
 muons in the shower N$_\mu$ to the number of electrons N$_e$, or the
 ratio of muon to electron densities at a certain distance from the
 shower axis.
 The type of the primary particles can only be studied in statistically
 big samples because of the fluctuations in the individual shower 
 development. Even then it is strongly affected by the differences
 in the hadronic interaction models.

 The optical method uses the fact that part of the particle ionization
 loss is in the form of visible light. All charged particles emit
 in air Cherenkov light in a narrow cone around their direction. 
 In addition to that charged particles excite Nitrogen atoms in the
 atmosphere, which emit fluorescence light. The output is not large,
 about 4 photons per meter, but the number of shower particles
 in UHECR showers is very large, and the shower can be seen from 
 distances exceeding 30 km. The fluorescence detection is very suitable
 for UHECR showers because the light is emitted isotropically and
 can be detected independently of the shower direction.
 Since optical detectors follow the shower track, the direction of 
 the primary cosmic ray is also relatively easy. The energy of the 
 primary particles is deduced from the total number of particles 
 in the shower development or from the number of particles at
 shower maximum. The rough number is that every particle at maximum
 carries about 1.5 GeV of primary energy. The mass of the primary
 cosmic ray nucleus is studied by the depth of shower maximum $X_{max}$,
 which is proportional to the logarithm of the primary energy per nucleon.

\subsection{The Highest Energy Cosmic Ray Event}
\begin{figure}[thb]
\includegraphics[width=0.48\textwidth]{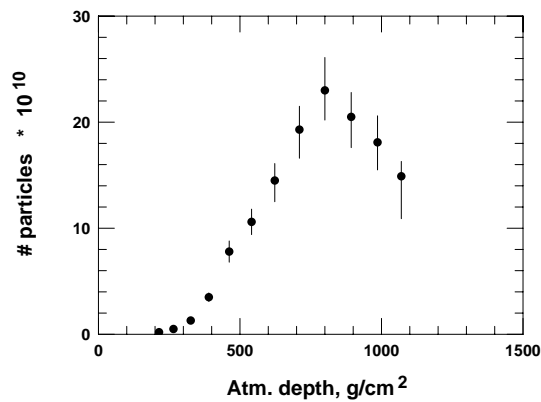}
\caption{The shower profile of the highest energy
 cosmic ray shower detected by the Fly's Eye.
}
\label{ts:fig1}
\end{figure}

 The highest energy cosmic ray particle was detected by the Fly's Eye 
 experiment~\cite{Bird_high}. We will briefly describe this event
 to give the reader an idea about these giant air showers. 
 The energy of this shower is estimated to be 3$\times$10$^{20}$ eV.
 This is an enormous macroscopic energy.
 3$\times$10$^{20}$ eV is equivalent to 4.8$\times$10$^{8}$ erg,
 7.2$\times$10$^{34}$ Hz or the energy of 290 km/h tennis ball
 in the favorite units of Alan Watson.
 Fly's Eye was the first air fluorescence experiment, located in
 the state of Utah, U.S.A. Fig.~\ref{ts:fig1} shows the shower 
 profile of this event as measured by the Fly's Eye.
 Note that the maximum of this shower contains more
 than 2$\times$10$^{11}$ electrons and positrons. Both the integral
 of this shower profile and the number of particles at maximum
 give about the same  energy.

 The errors of the estimates come from the errors of the individual
 data points, but mostly from the uncertainty in the distance 
 between the detector and the shower axis. The minimum energy of about
 2$\times$10$^{20}$ eV was calculated in the assumption that the
 shower axis was much closer to the detector than the data analysis
 derived. 

\section{ORIGIN OF UHECR}

  The first problem with the ultra high energy cosmic rays is
 that it is very difficult to imagine what their origin is. 
 We have a standard theory for the acceleration of cosmic rays 
 of energy below the knee of the cosmic ray spectrum at 
 galactic supernova remnants. This suggestion was first made
 by Ginzburg\&Syrovatskii in 1960's on the basis of energetics.
 The estimate was that a small fraction (5-10\%) of the kinetic
 energy of galactic supernova remnants is sufficient to maintain
 the energy carried by the galactic cosmic rays. The acceleration 
 process was assumed to be stochastic, Fermi type, acceleration
 that was later replaced with the more efficient acceleration at
 astrophysical shocks.

  This statement stands, but it is not applicable to all cosmic
 rays. Much more exact recent estimates and calculations show that
 the maximum energy achievable in acceleration on supernova
 remnant shocks is not higher than 10$^6$ GeV. This excludes
 not only UHECR, but also the higher energy galactic cosmic rays,
 that require supernova remnants in special environments~\cite{VB}.
 There are now some very interesting ideas about shock magnetic fields
 amplification by cosmic rays that lead to higher acceleration energy
 and flatter energy spectrum.
 See the talk of Pasquale Blasi for a discussion of new acceleration
 models.

\begin{figure}[thb]
\includegraphics[width=0.48\textwidth]{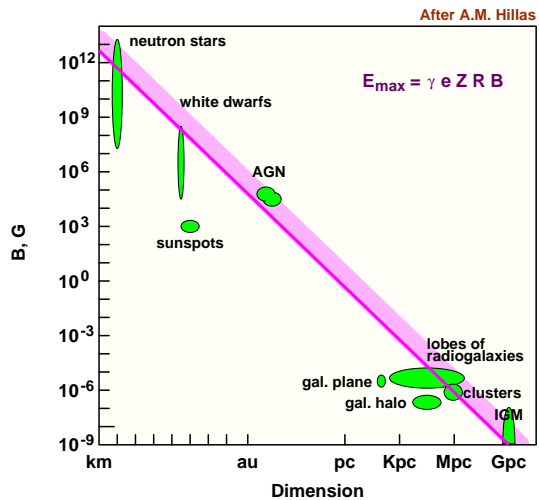}
\caption{Requirements for acceleration of charged nuclei at 
astrophysical objects as spelled out by A.M.~Hillas
(see text). 
}
\label{ts:fig2}
\end{figure}
 
  The reader should note that currently the acceleration of charged
 nuclei at supernova remnants is mostly a theoretical prediction.
 Supernova remnants have higher matter density than interstellar 
 space and one expects that the accelerated nuclei would interact 
 with the matter and generate high energy $\gamma$-rays. 

 Although Cherenkov gamma ray telescopes have observed
 supernova remnants with TeV $\gamma$-ray emission,
 there is no proof that TeV 
 and higher energy $\gamma$-rays are generated in hadronic 
 interactions. On the other hand, multi-wavelength observations
 show the existence of very strong shocks in supernova remnants
 with TeV $\gamma$-ray emission.

  We should then turn to extragalactic objects for acceleration to
 energies exceeding 10$^{20}$ eV. The scale for such acceleration was
 set up by Hillas~\cite{Hillas83} from basic dimensional 
 arguments. The first requirement for acceleration of charged nuclei
 in any type of object is that the magnetic field of the object
 contains the accelerated nucleus within the object itself. 
 One can thus calculate a maximum theoretical acceleration energy,
 that does not include and efficiency factor, as 
 $E_{max} \le \gamma e Z B R \; ,$
 where $\gamma$ is the Lorentz factor of the shock matter, $Z$ is the
 charge of the nucleus, $B$ is the magnetic field value. and $R$ is
 the linear dimension of the object. 

  Figure~\ref{ts:fig2}, which is a redrawn version of the original 
 figure of Hillas, shows what are the requirements for acceleration 
 to more than 10$^{20}$ eV. The lower edge of the shaded area shows
 the minimum magnetic field value for acceleration of iron nuclei
 as a function of the dimension of the astrophysical object.
 The upper edge is for acceleration of protons. 

 There are very few objects that can, even before an account for
 efficiency, reach that energy: highly magnetized neutron stars,
 active galactic nuclei,
 lobes of giant radiogalaxies, and possibly Gpc size shocks from
 structure formation. Other potential acceleration sites, gamma
 ray bursts, are not included in the figure because of the time
 dependence of magnetic field and dimension. 

 \subsection{Possible Astrophysical Sources of UHECR}

 In this subsection we give a brief description of some of the models 
 for UHECR acceleration at specific astrophysical objects. For a more
 complete discussion one should consult a review paper on
 the astrophysical origin of UHECR~\cite{ta04}, that contains an
 exhaustive list of references to particular models.
 
\begin{itemize}
\item {\bf Pulsars:} Young magnetized neutron stars with surface
 magnetic fields of 10$^{13}$ Gauss can accelerate charged iron nuclei
 up to energies of 10$^{20}$ eV~\cite{beo00}. The acceleration process
 is magnetohydrodynamic, rather than stochastic as it is at 
 astrophysical shocks. The acceleration spectrum is very flat proportional
 to 1/$E$. It is possible that a large fraction of the observed UHECR
 are accelerated in our own Galaxy. There are also models for UHECR
 acceleration at magnetars, neutron stars with surface magnetic fields
 up to 10$^{15}$ Gauss.

\item {\bf Active Galactic Nuclei:} As acceleration site of UHECR
 jets~\cite{HZ97} of AGN have the advantage that acceleration on the jet frame 
 could have maximum energy smaller than these of the observed UHECR
 by 1/$\Gamma$, the Lorentz factor of the jet. The main problem with
 such models is most probably the adiabatic deceleration of the particles
 when the jet velocity starts slowing down.

\item {\bf Gamma Ray Bursts:} GRBs are obviously the most energetic
 processes that we know about. The jet Lorentz factors needed to model
 the GRB emission are of order 100 to 1000. These models became popular
 with the realization that the arrival directions of the two most energetic
 cosmic rays coincide with the error circles of two powerful GRB.
 Different theories put the acceleration site at the inner~\cite{Wax95}
 or the outer~\cite{Vietri} GRB shock. To explain the observed UHECRs 
 with GRBs one needs fairly high current GRB activity, while most of
 the GRB with determined redshifts are at $Z\; >$ 1.

\item {\bf Giant Radio Galaxies:} One of the first concrete
 model for UHECR acceleration is that of Rachen\&Biermann, that
 dealt with acceleration at FR II galaxies~\cite{rb93}. Cosmic
 rays are accelerated at the `red spots', the termination shocks 
 of the jets that extend at more than 100 Kpc. The magnetic fields
 inside the red spots seem to be sufficient for acceleration up to
 10$^{20}$ eV, and the fact that these shocks are already inside 
 the extragalactic space and there will be no adiabatic 
 deceleration. Possible cosmologically nearby objects include 
 Cen A (distance of 5 Mpc) and M87 in the Virgo cluster (distance
 of 18 Mpc).

\item {\bf Quiet Black holes:} These are very massive quiet black
 holes, remnants of quasars, as acceleration sites~\cite{bg99}.
 Such remnants could be located as close as 50 Mpc from our Galaxy.
 These objects are not active at radio frequencies, but, if massive
 enough, could do the job. Acceleration to 10$^{20}$ requires a mass
 of 10$^9$ M$_\odot$.

\item {\bf Colliding Galaxies:} These systems are attractive with the
 numerous shocks and magnetic fields of order 20 $\mu$G that have 
 been observed in them~\cite{CJC92}. The sizes of the colliding galaxies
 are very
 different and with the observed high fields may exceed the
 gyroradius of the accelerated cosmic ray.

\item {\bf Clusters of Galaxies:} Magnetic fields of order several
 $\mu$G have been observed at lengthscales of 500 Kpc. Acceleration
 to almost 10$^{20}$ eV is possible, but most of the lower energy
 cosmic rays will be contained in the cluster forever and only
 the highest energy particles will be able to escape~\cite{krj98}.
  
\item {\bf Gpc scale shocks from structure formation:} A combination
 of Gpc scales with 1 nG magnetic field satisfies the Hillas 
 criterion, however the acceleration at such shocks could be much
 too slow, and subject to large energy loss.
\end{itemize}  
 
\subsection{Top-Down Scenarios}

 Since it became obvious that the astrophysical acceleration up to
 10$^{20}$ eV and beyond is very difficult and unlikely, a large 
 number of particle physics scenarios were discussed as explanations
 of the origin of UHECR. To distinguish them from the acceleration
 ({\em bottom-up}) processes they were called {\em top-down}.
 The basic idea is that very massive (GUT scale) $X$ particles decay 
 and the resulting fragmentation process downgrades the energy 
 to generate the observed UHECR. Since the observed cosmic rays 
 have energies orders of magnitude lower than the $X$ particle mass,
 there are no problems with achieving the necessary energy scale.
 The energy content of UHECR is not very high, and the $X$ particles 
 do not have to be a large fraction of the dark matter.
 There is a large number of topological defect models which are
 extensively reviewed in Ref.~\cite{bs00}.
 
 There are two distinct branches of such theories. One of them
 involves the emission of $X$ particles by topological defects.
 This type of models follows the early work of C.T.~Hill~\cite{hill83}
 who described the emission from annihilating monopole/antimonopole
 pair, which forms a metastable monopolonium. The emission of 
 massive $X$ particles is possible by superconducting cosmic string
 loops as well as from cusp evaporation in normal cosmic strings and
 from  intersecting cosmic strings. The $X$ particles then decay in
 quarks and leptons. The quarks hadronize in baryons and mesons,
 that decay themselves along their decay chains. The end result 
 is a number of nucleons, and much greater (about a factor of 30
 in different hadronization models) and approximately equal 
 number of $\gamma$-rays and neutrinos.   
   
 A monopole is about 40 heavier than a $X$ particle, so every monolonium
 can emit 80 of them. Using that number one can estimate the number
 of annihilations that can provide the measured UHECR flux, which turns
 out to be less than 1 per year per volume such as that of the Galaxy.
 Another possibility is the emission of $X$ particles from cosmic
 necklaces - a closed loop of cosmic string including monopoles. 
 This particular type of topological defect has been extensively
 studied~\cite{bv97}.
 
 The other option is that the $X$ particles themselves are remnants 
 of the early Universe. Their lifetime should be very long, 
 maybe longer than the age of the Universe~\cite{bkv97}. They could
 also be a significant part of the cold dark matter. Being superheavy,
 these particle would be gravitationally attracted to the Galaxy 
 and to the local supercluster, where their density could well exceed
 the average density in the Universe. 
  
 There are two main differences between bottom-up and top-down models
 of UHECR origin. The astrophysical acceleration generates charged 
 nuclei, while the top-down models generate mostly neutrinos and
 $\gamma$-rays plus a relatively small number of protons. 
 The energy spectrum of the cosmic rays that are generated in the
 decay of $X$ particles is relatively flat, close to a power law
 spectrum of index $\alpha$=1.5. The standard acceleration energy
 spectrum has index equal to or exceeding 2.

\subsection{Hybrid Models}

 There also models that are hybrid, they include elements of both groups.
 The most successful of those is the Z-burst model~\cite{tjw99,fms99}.
  The idea is that somewhere in the Universe neutrinos of ultrahigh
 energy are generated. These neutrinos annihilate with cosmological
 neutrinos in our neighborhood and generate $Z_0$ bosons which decay
 and generate a local flux of nucleons, pions, photons and neutrinos.
 The resonant energy for $Z_0$ production is
 4$\times$10$^{21}$ eV/$m_\nu$(eV), where $m_\nu$ is the mass of the
 cosmological neutrinos. The higher the mass of the 
 cosmological neutrinos is, the lower the resonance energy requirement.
 In addition the cosmological neutrinos are gravitationally attracted 
 to concentrations of matter and their density increases in our
 cosmological neighborhood. If the neutrino masses are low, of order
 of the mass differences derived from neutrino oscillations, the
 energy of the high energy neutrinos should increase.

\section{PROPAGATION OF UHECR}

 Particles of energy 10$^{20}$ eV can interact on almost any target.
 The most common, and better known, target is MBR. It fills the whole
 Universe and its number
 density of 400 cm$^{-3}$ is large. The interactions on the 
 radio and infrared backgrounds (IRB) are also important. Let us have
 a look at the main processes that cause energy loss of nuclei
 and gamma rays.

\subsection{Energy Loss Processes}

 The main energy loss process for protons is the photoproduction on
 astrophysical photon fields $p\gamma \rightarrow p + n\pi$.
 The minimum center of mass energy for photoproduction is
 $\sqrt {s_{thr}} = m_p + m_{\pi^0}\; \sim$ 1.08 GeV.
 Since $s = m_p^2 + 2 (1 - \cos{\theta}) E_p \epsilon$
 (where $\theta$ is the angle between the two particles) one can
 estimate the proton threshold energy for photoproduction on the MBR
 (average energy $\epsilon\; =$ 6.3$\times$10$^{-4}$ eV).
 For $\cos{\theta}$ = 0 the proton threshold energy is
 $E_{thr}$ =  2.3$\times$10$^{20}$ eV. Because there are head to head
 collisions and because the tail of the MBR energy spectrum continues
 to higher energy, the intersection cross section is non zero above
 proton energy of 3$\times$10$^{19}$ eV.

\begin{figure}[thb]
\includegraphics[width=0.48\textwidth]{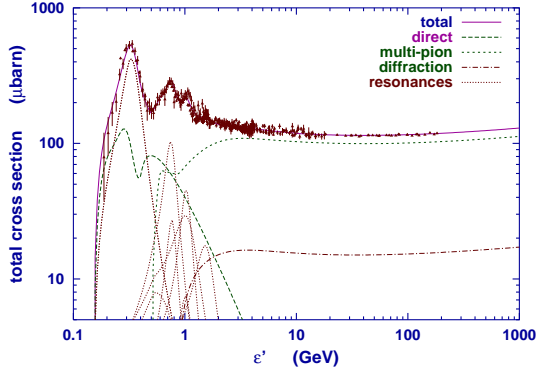}
\caption{Photoproduction cross section as a function of the
photon energy for stationary proton targets. 
}
\label{ts:fig3}
\end{figure}

 The photoproduction cross section is very well studied in accelerator
 experiments and is  known in detail. Figure~\ref{ts:fig3} shows the
 photoproduction cross section in the mirror system~\cite{SOPHIA},
 as a function of
 the photon energy  for stationary protons, i.e. as it is measured in
 accelerators. At threshold the most important process is the 
 $\Delta^+$ production where the cross section reaches a peak exceeding
 500 $\mu$b. It is followed by a complicated range that includes the 
 higher mass resonances and comes down to about 100 $\mu$b. After that
 one observes an increase that makes the photoproduction cross
 section parallel to the $pp$ inelastic cross section. The neutron 
 photoproduction cross section is nearly identical.

 Another important parameter is the proton inelasticity $k_{inel}$,
 the fraction of its energy that a proton loses in one interaction.
 This quantity is energy dependent. At threshold protons lose
 about 18\% on their energy. With increase of the CM energy this
 fractional energy loss increases to reach asymptotically 50\%.

\begin{figure}[thb]
\includegraphics[width=0.48\textwidth]{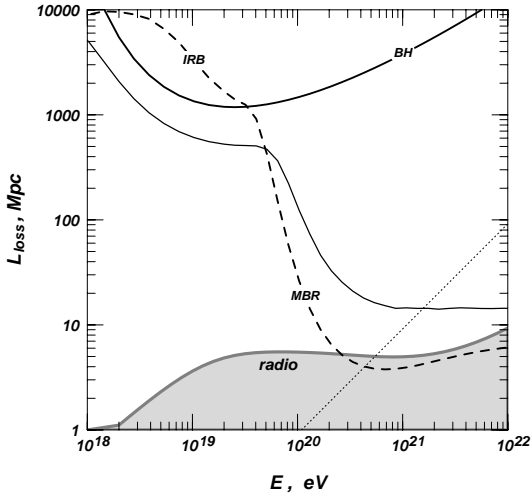}
\caption{Energy loss length of protons in interactions in MBR 
 and IRB. The shaded area shows the \protect$\gamma\gamma$
 interaction length.
}
\label{ts:fig4}
\end{figure}

 The proton pair production $p\gamma \rightarrow e^+e^-$ is the
 same process that all charged particles suffer in nuclear fields.
 The cross section is high, but the proton energy loss is of order
 $m_e/m_p$ $\simeq$ 4$\times$10$^{-4}$E. Figure~\ref{ts:fig4} shows the 
 energy loss length $L_{loss} = \lambda/k_{inel}$
 (the ratio of the interaction length to the inelasticity coefficient)
 of protons in interactions in the microwave and infrared backgrounds. 
 
 The dashed line shows the proton interaction length and one can see 
 the increase of $k_{inel}$ in the ratio of the interaction length to
 energy loss length. The contribution of the pair production is shown
 with a thin line. The energy loss length never exceeds 4,000 Mpc,
 which is the adiabatic energy loss due to the expansion of the 
 Universe for $H_0$ = 75 km/s/Mpc. The dotted line shows the neutron
 decay length. Neutrons of energy less than about 3$\times$10$^{20}$ eV 
 always decay and only higher energy neutrons interact.

\begin{figure}[thb]
\includegraphics[width=0.48\textwidth]{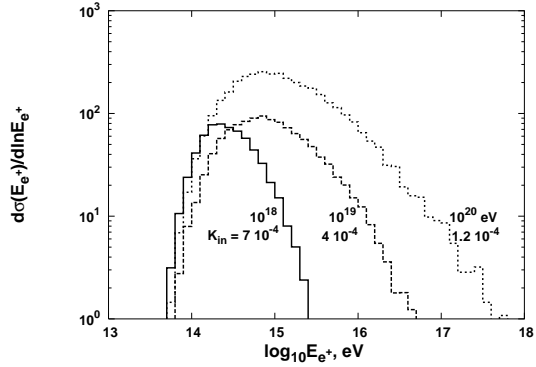}
\caption{Energy distribution of positrons generated in pair 
production interactions of protons of fixed energy in the MBR.
}
\label{ts:fig5}
\end{figure}

 The pair production process deserves more attention since it will
 become important soon. Figure~\ref{ts:fig5} shows the positron 
 spectra produced in pair production interactions of protons with
 fixed energy. Next to the proton energy the figure indicates the
 inelasticity coefficients in these interactions.
 The pair production cross section grows with the proton energy while
 the inelasticity coefficient decreases. The combination of both 
 these parameters generates the energy loss length that has a minimum
 at about 2$\times$10$^{19}$ eV.

\begin{figure}[thb]
\includegraphics[width=0.48\textwidth]{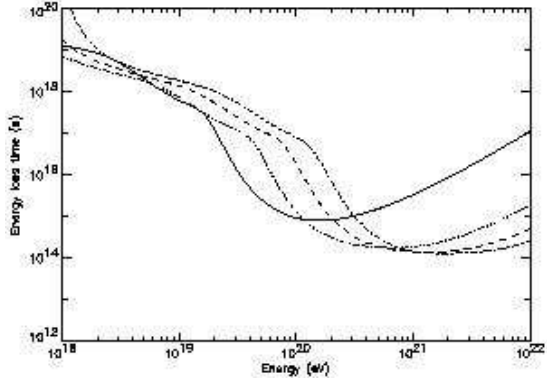}
\caption{Energy loss length of heavy nuclei (He, O, Si, Fe from
 left to right in MBR~\protect\cite{Bertone}.
}
\label{ts:fig7}
\end{figure}

 Heavier nuclei lose energy to a different process - photodisintegration,
 loss of nucleons mostly at the giant dipole resonance~\cite{PSB76}.
 Since the  relevant energy in the nuclear frame is of order 20 MeV,
 the process starts at lower energy. The resulting nuclear fragment may
 not be stable. It then decays and speeds up the energy loss of the whole
 nucleus.
 Ultra high energy heavy nuclei, where the energy per nucleon is higher
 than photoproduction, have also loss on photoproduction. The energy loss 
 length for He nuclei in photodisintegration is as low as 10 Mpc at
 energy of 10$^{20}$ eV.
 Heavier nuclei reach that distance at higher total energy. 
 Figure~\ref{ts:fig7} shows the energy loss time of heavy nuclei -
 10$^{14}$ s equals approximately 1 Mpc. 

 UHE gamma rays also interact on the microwave background. The main
 process is $\gamma \gamma \rightarrow e^+ e^-$. This is a resonant 
 process and for interactions in the MBR the minimum interaction 
 length is achieved at 10$^{15}$ eV. The interaction length in MBR
 decreases at higher $\gamma$-ray energy and would be about a 50 Mpc
 at 10$^{20}$ eV if not for the radio background. The radio background
 does exist but its number density is not well known. Figure~\ref{ts:fig4}
 shows the interaction length for this process in the MBR and
 the radio background as a shaded area.  

 The fate of the electrons produced in a $\gamma \gamma$ collision 
 depends on the strength of the magnetic fields in which UHE electrons
 lose energy very fast. The photon energy is than quickly downgraded
 and the $\gamma \gamma$ interaction length becomes very close
 to the gamma ray energy loss length. In the case of very low magnetic
 fields (0.01 nG) the synchrotron energy loss is low (it is proportional
 to $E_e^2 B^2$) and then inverse Compton scattering (with a cross section
 very similar to this of $\gamma \gamma$) and cascading is possible.
 The energy loss length of the gamma rays would be
 higher in such a case.

 The general conclusion from this analysis of the energy loss of
 protons and gamma rays in their propagation through the Universe
 is these UHE particles can not survive at distances of more than
 few tens of Mpc and sources of the detected cosmic rays have to
 be located in our cosmological neighborhood. Every small increase
 of the distance between the source and the observer would require
 increase of the maximum energy at acceleration (or other
 production mechanism) and will affect significantly the 
 energy requirement for the UHECR sources.

\subsection{Modification of the Proton Spectrum in Propagation.
 Numerical Derivation\\ of the GZK Effect.}

\begin{figure}[thb]
\includegraphics[width=0.48\textwidth]{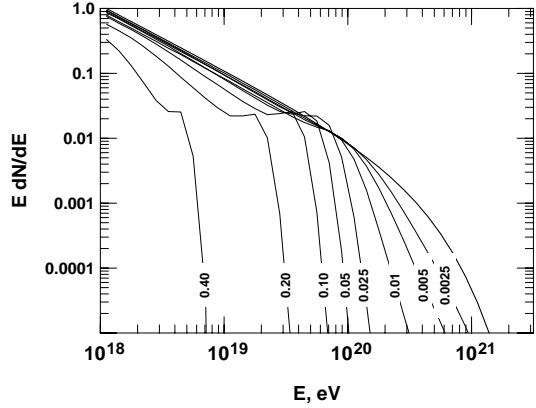}
\caption{ Evolution of the cosmic ray spectrum 
 in propagation through different redshifts.
}
\label{ts:fig8}
\end{figure}

 Figure~\ref{ts:fig8} shows the evolution of the spectrum of protons 
 because of energy loss during propagation at different redshifts.
 The thick solid lines shows the spectrum injected in intergalactic
 space by the source, which in this exercise is 
$$ \frac{dN}{dE} \, = \, A \times E^{-2}/ \exp({E/10^{21.5}}
 {\rm eV}) \; .$$
 After propagation on 10 Mpc ({\em z} = 0.0025) some of the highest
 energy protons are missing. This trend continues with distance and
 at about 40 Mpc  another trend appears - the flux of protons of
 energy just below 10$^{20}$ eV
 is above the injected one. This is the beginning of the formation
 of a pile-up in the range where the photoproduction cross section
 starts decreasing. Higher energy particles that are downgraded in 
 this region lose energy less frequently and a pile-up is developed.

 The pile-up is better visible in the spectra of protons propagated at 
 larger distances. One should remark that the size of the pile-up 
 depends very strongly on the shape of the injected spectrum. If it
 had a spectral index of 3 instead of 2 the size of the pile-up 
 would have been barely visible as the number of high energy particles
 decreased by a factor of 10.

 When the propagation distance exceeds redshift of 0.4 there are no more 
 particles of energy above 10$^{19}$ eV independently of the
 maximum acceleration energy. All these particles have lost
 energy in photoproduction, pair production and adiabatic losses.
 From there on most of the losses are adiabatic. 

 In order to obtain the proton spectrum created by homogeneously and
 isotropically distributed cosmic ray sources filling the whole
 Universe one has to integrate a set of such (propagated) spectra
 in redshift using the cosmological evolution of the cosmic ray
 sources, which is usually assumed to be the same as that of the star
 forming regions (SFR) $\eta(z) \; = \; \eta(0)(1 + z)^n$ with 
 $n$ = 3, or 4 up to the epoch of maximum activity $z_{max}$ and 
 then either constant or declining at higher redshift. High redshifts
 do not contribute anything to UHECR ($z$ = 0.4 corresponds to a
 propagation distance of 1.6 Gpc for $H_0$ = 75 km/s/Mpc).
 After accounting for the increased
 source activity the size of the pile-ups has a slight increase.
  
  Apart from the pile-up, there is also a dip at about
 10$^{19}$ eV which is due to the energy loss on pair production.
 It is also preceded by a small pile-up at the transition from
 adiabatic to pair production loss.
 This feature was first pointed at by Berezinsky\&Grigorieva~\cite{BerGri}.

 The GZK cutoff, the pile-up and the pair production dip 
 characterize the energy spectrum of extragalactic protons under
 the assumptions of injection spectrum shape, cosmic ray luminosity,
 cosmological evolution and isotropic distribution of the cosmic
 ray sources in the Universe.

\subsection{Modification of the Gamma Ray Spectra from {\em top-down}
 models.}

\begin{figure}[thb]
\includegraphics[width=0.48\textwidth]{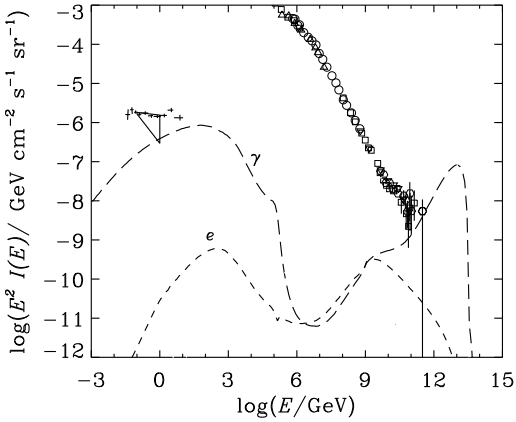}
\caption{ Evolution of the energy spectra of electrons and
 \protect$\gamma$-rays injected in a top-down scenario with
 \protect$X$ particle mass of 10\protect$^{14}$ GeV.
}
\label{proth_sta}
\end{figure}

  Because of the strong influence of the radio background and of the
 cosmic magnetic fields the modification of the spectrum of gamma 
 rays in a top-down scenario is much more difficult to calculate
 exactly. There are, however, many general features that are common 
 in any of the calculations. Figure~\ref{proth_sta} shows the
 gamma ray spectrum emitted in a top-down scenario with $m_X$ =
 10$^{14}$ GeV~\cite{proth_sta}.
 The spectra of $\gamma$-rays and electrons from the $X$ 
 decay chain are indicated with different line types.

 The typical gamma ray energy spectrum in {\em top-down} models is E$^{-3/2}$.
 We shall start the discussion from the highest energy and follow 
 the energy dissipation in propagation.  
 The highest energy gamma rays have not suffered significant losses.
 At slightly lower energy, though, the $\gamma \gamma $ cross section
 grows and the energy loss increases. One can see the dip at about
 10$^{10 - 11}$ GeV which is caused by the radio background.
 The magnetic field is assumed
 sufficiently high that all electrons above 10$^9$ GeV immediately
 lose energy in synchrotron radiation. 
 The minimum ratio of the gamma-ray to cosmic ray flux 
 is reached at about 10$^{15}$ eV, where the minimum of the
 interaction length in the MBR is, after 
 which there is some recovery. There is another absorption
 feature from interactions on the infrared background.
 The gamma ray peak in the GeV region consists mostly of synchrotron photons.
 Isotropic GeV gamma rays, that have been measured, can be used to
 restrict top-down models in some assumptions for the magnetic field
 strength. 

\subsection{UHECR Propagation and Magnetic Fields}

 The possible existence of non negligible extragalactic magnetic fields
 would modify the propagation of the UHE cosmic rays
 independently of their nature and origin. There is little observational
 data on these fields. The estimate of the average strength of
 these fields in the Universe is 10$^{-9}$ Gauss (1 nG)~\cite{Kronberg94}.
 On the other hand $\mu$G fields have been observed in clusters of 
 galaxies, and in a bridge between two parts of the Coma cluster.
  
 Even fields with nG strength would affect the propagation of
 UHE cosmic rays. If UHECR are protons or heavier nuclei
 they would scatter of these fields.
 This scattering would lead to deviations from the source direction and 
 to an increase of the pathlength from the source to the observer.
 It would make the source directions less obvious and would create a
 magnetic horizon~\cite{Stanevetal01} for extragalactic protons of
 energy below 10$^{19}$ eV as their propagation time from the source
 to the observer would  start exceeding Hubble time. When the horizon
 is achieved the cosmic ray spectrum appears flatter than it actually is.

\begin{figure}[thb]
\includegraphics[width=0.48\textwidth]{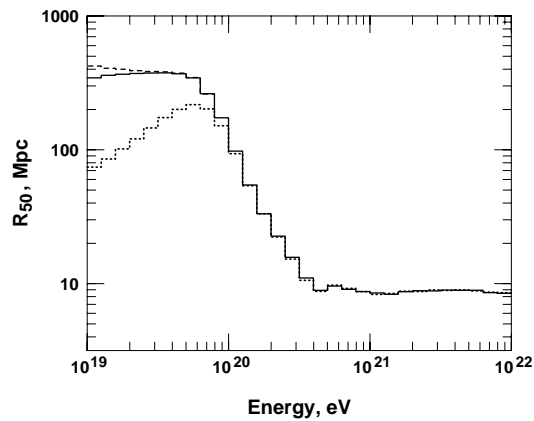}
\caption{Distance from the source that contains more than 50\%
 of the emitted protons.
}
\label{horizon}
\end{figure}

 If regular magnetic fields of strength exceeding 10 nG were present on
 10 Mpc coherence length they would lead to significant biases in the 
 propagated spectra~\cite{SSE03} in a function of the relative geometry
 of the field, source and observer. 10$^{19}$ eV particles would
 gyrate around the magnetic field lines and thus appear coming from 
 a wide range of directions. The flux of such particles would be much
 higher than at injection. Only particles of energy above 10$^{20}$ eV
 would be able to propagate through the magnetic field lines.

  Ultrahigh energy protons are also affected in propagation by
 the galactic magnetic fields - they scatter and acquire an
 angle with their direction outside of the Galaxy. That angle
 depends on the UHECR rigidity and direction. The largest angle 
 should be when the proton has to propagate close to the galactic
 center and galactic bulge region where the fields, although
 not exactly known, are the highest. Excluding the galactic center 
 vicinity, the average deflection angle for 10$^{20}$ eV protons
 is between 3.1$^o$ and 4.5$^o$ in different galactic magnetic 
 field models~\cite{stanev97}. 

\subsection{Production of Secondary Particles in Propagation}

 One interesting feature that can be used for testing of the type
 and distribution of UHECR sources is the production of secondary
 particles in propagation. The energy loss of the primary protons
 and $\gamma$-rays is converted to secondary gamma rays and neutrinos
 (in the case of primary nuclei). Gamma rays are generated in
 nucleon photoproduction interactions and in BH pair production
 processes as well as in $\gamma\gamma$ collisions.In the case of
 isotropic and homogeneous source distribution the gamma ray 
 energy is degraded and eventually converted to MeV/GeV diffuse
 isotropic flux. The value of this flux could be used to restrict
 the amount of energy in UHECR~\cite{BerSmi,proth_sta}. Some
 cosmologically nearby sources may still create a halo in the
 source direction of high energy gamma rays that could be detected 
 by the sensitive contemporary TeV gamma ray telescopes.

\begin{figure}[thb]
\includegraphics[width=0.48\textwidth]{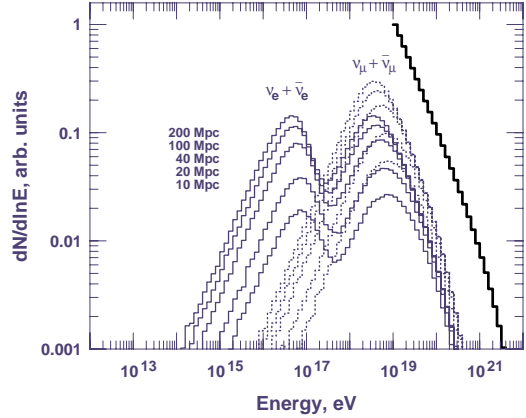}
\caption{ Production of neutrinos from cosmic ray proton
propagation on different distances from 10 to 200 Mpc 
}
\label{propneu}
\end{figure}

 Most interesting are the cosmogenic neutrinos, that were first 
 proposed by Berezinsky \& Zatsepin~\cite{BZ68} and have been since
 calculated many times, most recently in Ref.~\cite{ESS01}.
 Every charged pion produced in a photoproduction interaction
 generates three neutrinos through its decay chain. 

 The spectrum of cosmogenic neutrinos depends on the UHECR 
 spectrum, the UHECR source distribution and very strongly
 on the cosmological evolution of the UHECR sources~\cite{SS05}.
 The sensitivity to the cosmological evolution of the sources
 is very high because of the lack of energy loss (except for
 adiabatic loss) of the generated neutrinos. Figure~\ref{propneu}
 shows the spectra of neutrinos generated in proton propagation
 at different distances. In the contemporary Universe muon neutrinos
 and antineutrinos peak at about 10$^{18}$ eV, while electron
 neutrinos and antineutrinos show a double peaked spectrum.
 The higher energy peak, coinciding with that of $\nu_\mu$
 consists mostly of electron neutrinos, while the lower 
 energy one is of $\bar{\nu}_e$ from neutron decay. If
 there is a high fraction of heavy nuclei in the primary
 UHECR the $\bar{\nu}_e$ would dominate the $\nu_\mu$ flux
 since there would be many more neutrons from photodisintegration
 than photoproduction interactions, which mostly protons
 and He nuclei would suffer.

\begin{figure}[thb]
\includegraphics[width=0.48\textwidth]{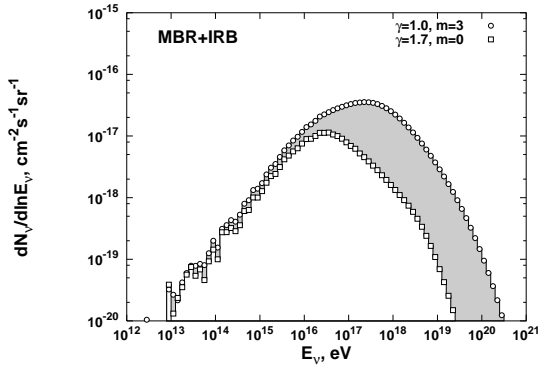}
\caption{Cosmogenic \protect$\nu_\mu + \bar{\nu}_\mu$ generated
 by protons with a flat acceleration spectrum ($\gamma$=1.0) with 
 cosmological evolution and a steep one ($\gamma$=1.7) without
 evolution.  
}
\label{allnu}
\end{figure}

 It is of some importance to note that MBR is not the only 
 target for neutrino production. The second most important one 
 is the isotropic infrared and optical background (IRB). 
 Its number density is, or course, much lower, but lower 
 energy protons can interact in it and even in the case of
 flat acceleration spectra the number of interacting protons
 to a large extent compensates for the lower photon target 
 density.

 Fig.~\ref{allnu} shows the spectra of $\nu_\mu$+$\bar{\nu}_\mu$ 
 generated by a flat ($\gamma$=1.0) and steep ($\gamma$=1.7)
 UHECR acceleration spectra. In the case we are lucky enough 
 to detect cosmogenic neutrinos with the neutrino telescopes 
 under construction and design they could help a lot in 
 limiting the models for the origin of the ultrahigh energy
 cosmic rays.

\section{Experimental data}  

  At this meeting we saw the first set of 
 experimental data with very good statistics observed by
 the Auger Collaboration. Auger reported the data from an 
 exposure of more than 5,165 km$^2$.ster.yr obtained during 
 the construction of the array. This exceeds the exposure 
 of the previous biggest array, Agasa~\cite{AGASA}, by about
 a factor of three. The exposure of the HiRes detector is 
 energy dependent as the fluorescent light of higher energy
 showers can be detected from larger distances.
 The Auger air shower array consists of
 1,600 water tanks of area 10 m$^2$ in which the charged
 shower particles and the converted $\gamma$-rays produce
 Cherenkov light. Tanks are viewed with 3 photomultipliers.
 The water tanks have the advantage to have significant 
 effective area for highly inclined showers.

\subsection{Cosmic ray spectrum}

 Auger reported three energy spectra obtained in different manner:
 one from the surface array normalized to the fluorescent 
 telescopes measurement~\cite{mroth}, another from hybrid array,
 i.e. from showers observed
 both by the ground array and by the fluorescent
 detectors~\cite{lperrone}, and
 a third one coming from showers arriving at zenith angles exceeding
 60$^\circ$~\cite{pfacal}. All three spectra agree with each other
 within the  statistical uncertainties. There is only two events
 of energy above 10$^{20}$ eV in this set. The presented spectra
 thus support the conclusion of HiRes~\cite{HiRes1,HiRes2} that the cosmic
 ray spectrum does not continue above 10$^{20}$ eV with the same
 $\sim$E$^{-2.7}$ spectral index as the Agasa experiment found.
 There is obviously a steepening
 of the spectrum which may be  consistent with a GZK feature.

 There are good reasons to trust the spectrum measurements of the
 Auger collaboration. The analysis only includes well contained
 showers by the requirement that the highest hit detector is 
 surrounded by six active detectors. This requirement also 
 guarantees that there is enough information for a good shower
 analysis. Another requirement is the reconstructed shower core
 position is inside the 3,000 km$^2$ array. For this reason the
 exposure of Auger is very well known.
 Only events with reconstructed energy above 3$\times$10$^{18}$ eV,
 where the efficiency is 100\%, are included.
 The uncertainty on the energy estimate
 is quoted to be 22\% - most of it due to fluorescent efficiency.
\begin{figure}[thb]
\includegraphics[width=0.48\textwidth]{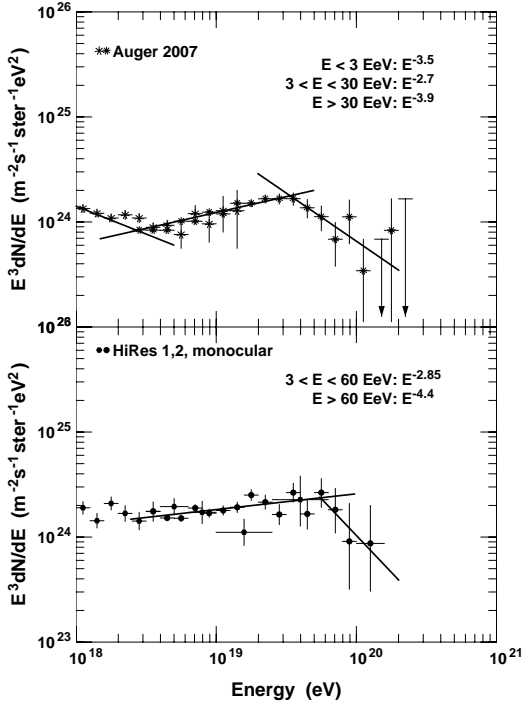}
\caption{Comparison of the detailed spectral shapes reported 
 by the Auger Collaboration (surface detector and hybrid events)
 and by HiRes 1 and 2 in monocular mode.  
}
\label{AugerHR}
\end{figure}

 The overall normalization of the spectrum is somewhat lower than 
 that of HiRes. It also seems to have slightly different shape:
 the dip in the spectrum above 10$^{19}$ eV (seen mostly in the
 {\em hybrid} data set appears deeper and the recovery faster
 as shown in Fig.~\ref{AugerHR}. The depletion of high energy
 showers is not that steep and starts a bit earlier. Note that
 the slopes presented in Fig.~\ref{AugerHR} are not the same
 as presented by Auger at the meeting~\cite{tyamamoto} - my fits
 are probably not as careful as those of the group. 
 As small these differences are, they currently affect the 
 comparison with the UHECR acceleration and propagation models
 and the derivation of the end of the galactic cosmic ray 
 spectrum (see the talk of Venya Berezinsky in this volume).
 The solution of these questions will not be made before we
 have a very good measurement of the UHECR chemical composition.

\subsection{UHECR composition}

\begin{figure}[thb]
\includegraphics[width=0.48\textwidth]{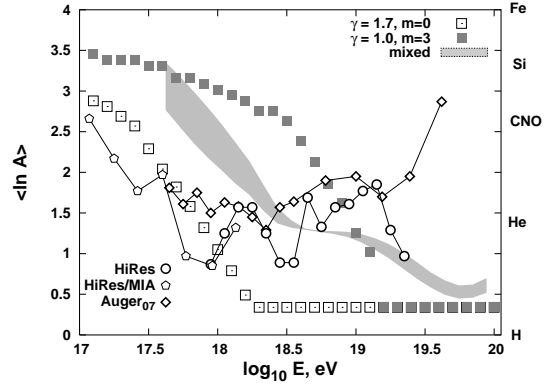}
\caption{The data of HiRes/MIA, HiRes and Auger are 
 compared to the predictions of three models of UHECR
 acceleration and propagation.  
}
\label{compos}
\end{figure}

 Auger also presented data on the average depth of shower 
 maximum ($X_{max}$) as a function of the shower energy~\cite{munger}.
 The average $X_{max}$ is the measure of the cosmic ray 
 chemical composition that could be made by fluorescent 
 detectors. Hybrid events are used in this data set because
 even one surface detector triggered in coincidence with
 the fluorescent trigger vastly improves the shower
 reconstruction. Fig.~\ref{compos} compares the measurements
 of HiRes, HiRes Prototype/MIA and Auger, converted to
 $\langle lnA \rangle$ using the Sibyll2.1 hadronic interaction model.
 The use of other models would change the derived values, i.e.
 move the experimental points to lower $\langle lnA \rangle$ values for 
 models having shallower $X_{max}$.
 The two sets of squares and the shaded area plot roughly the
 $\langle lnA \rangle$ values that come from three types of models: the empty
 squares correspond to the model of Berezinsky et al~\cite{Ber1,Ber2},
 full ones correspond to the model of Bahcall \& Waxman~\cite{BW}
 and the shaded area corresponds to the mixed primary composition
 model~\cite{Allard1,Allard2}. At very high energy in Fig.~\ref{compos}
 I have assumed a standard H + He composition in a ratio of 9:1.

 The model of Berezinsky {\em et al} is that all cosmic rays above
 10$^{18}$ eV are of extragalactic origin, and are mostly protons.
 The acceleration (injection) spectrum is steep with a slope of 1.7.
 The dip is due to pair production losses. The shape of the spectrum
 below the dip is determined by the transition to purely adiabatic
 energy loss. There is no need for a strong cosmological evolution
 of the cosmic ray sources. The maximum amount of He and heavier
 nuclei is about 15\%.
  Bahcall \& Waxman assume that the dip is 
 where the flat extragalactic UHECR flux ($\gamma$=1) intersects the
 flux of galactic cosmic rays. A similar model was also proposed by
 Wibig \& Wolfendale~\cite{WW} with slightly different parameters.
 The flat acceleration spectrum models require strong cosmological
 evolution $(1 + z)^{3-4}$ to be able to fit the experimental data. 
 Since the extragalactic cosmic rays are mostly H the composition becomes
 very light after the transition. Obviously in the Berezinsky {\em et al} 
 model there is no need of galactic cosmic rays of energy above
 10$^{18}$ eV, while in the other models the Galaxy has to accelerate
 particles to energy higher by an order of magnitude.

 Mixed UHECR composition model assumes that these particles
 leave their sources with a composition similar to those of GeV 
 galactic cosmic rays. The composition changes in propagation and
 above 10$^{19.5}$ eV the composition also becomes very light.
 The injection spectrum is relatively flat ($\gamma$=1.2-1.3).
 Since the spectrum at Earth depends on the primary composition 
 as well as on the acceleration spectrum it is possible to fit
 the observations in different ways. The current experimental data 
 on the cosmic ray composition do not seem to support any of the
 models. 

 There is, however, a problem in our understanding of the shower
 development in the atmosphere. If Auger surface detector data were
 analyzed using existing shower simulations (not normalized to the
 fluorescent detector) the energy estimate would on average go up by
 20-25\% and the composition would appear heavier~\cite{rengel}.
 The only current hadronic interaction model that predicts similar
 $X_{max}$  and high surface (muon) density is EPOS~\cite{EPOS},
 which is not yet well studied. 

 For this reason the solution of the problems of the highest energy
 cosmic rays will have to wait. The differences in the Auger 
 energy estimates by the surface and fluorescent detectors is very
 similar to the current differences between the AGASA and HiRes
 spectra after the energy estimate of AGASA has come down by 10-15\%
 with the use of the contemporary hadronic interaction model~\cite{Tesh06}.
 The hybrid detection of Auger and the Telescope Array,
 as well as the theoretical work, will lead us to the solution in 
 the next several years. 

 One of these problems is almost solved: many experiments and currently
 Auger~\cite{Aug_gam} have set strict limits on the fraction of gamma-rays
 in the UHECR flux. For showers above 10$^{19}$ eV the limit of Auger
 is 2\%. At higher energy the limits are not that strict because of 
 limited statistics. A general conclusion can be drawn from these limits
 that UHECR are not the result of {\em top-down} models and are due
 to acceleration in powerful astrophysical objects.

\subsection{Arrival directions of UHECR}

 The question then is which these objects are. The AGASA
 experiment~\cite{agasadir} has seen clustering of the 
 highest energy events, i.e. several groups of two events 
 and one of three events that come from similar directions
 smaller than the angular resolution of the array. HiRes
 did not observe that type of clustering. Auger~\cite{smollerach}
 does not confirm the clustering either, although there is still
 some possibility that some degree of clustering
 (2\% probability for isotropic distribution) exists among 
 the highest energy events.  
 
 So we are back to that controversial situation where we know
 that the astrophysical sources of UHECR have to be close to us
 in cosmological sense but we can not see them. The Auger 
 Collaboration is in the process of an intensive search 
 for anisotropies and correlations with different types of
 astrophysical objects in their data sample. Since we expect 
 a very strong increase in their statistics even during the 
 next year we hope future data will help resolving this 
 situation.

 \section{Summary}

 The cosmic ray spectrum steepens at the approach of 10$^{20}$ 
 eV according to the HiRes and the new Auger data. Is this 
 steepening the long expected GZK feature or is it of a different
 origin? An example for a different origin could be the 
 inability of the sources to accelerate cosmic rays of
 energy higher than 5$\times$10$^{19}$ eV. Solving this would
 require higher statistics than is currently available.

 The current statistics is, however, enough to establish
 the fact that 1-2$\times$10$^{19}$ eV cosmic rays are not $\gamma$-rays.
 This fact supports the acceleration {\em bottom-up} scenarios
 for cosmic ray production. At the higher energies more
 statistics is needed to set limits of a higher quality.
 Thus it is still possible (although not likely) that the
 highest energy events could result from {\em top-down} models.
 
 It is not currently obvious what the cosmic ray nuclear 
 composition is in the energy range above 10$^{18}$ eV. The Auger
 data points at a composition that is somewhat heavier than
 the one derived by the HiRes experiment. Differences are not
 statically significant yet. Composition measurements are hurt
 by the insufficient understanding of the shower development.
 Current hadronic interaction models should improve after 
 comparisons with the Large Hadronic Collider (LHC) data that
 will become available soon.

 We still can not see the UHECR sources. Studies of anisotropy,
 in addition of revealing the UHECR sources, are complimentary
 to the composition studies. If these particles were heavy 
 nuclei they would scatter stronger in the magnetic fields and
 show smaller correlation with their sources. Protons would more
 or less point at their sources. The identification of the UHECR
 sources would also contribute to our knowledge of magnetic fields
 in the Galaxy and the Universe.

 We are now in a period when the UHECR statistics is increasing 
 very fast and will help the solution of the problems listed above.
 In addition to Auger and TA the work on a satellite based 
 UHECR Observatory continues. While the previous two projects,
 EUSO~\cite{EUSO} and OWL~\cite{OWL} barely exist anymore,
 the Japanese JEM/EUSO project, that is to be installed at
 the Japanese module of the International Space Station is 
 doing well and may be launched in 2013. Such an experiment would 
 increase the statistics by another factor of 10 over the surface
 arrays.

 Finally, the neutrino astronomy projects are also in the 
 process of fast development. Auger is looking for horizontal
 air showers initiated by neutrinos, the construction of
 IceCube~\cite{IceCube} is going extremely well, and the
 radio detection of UHE neutrinos is also moving forward.
 The possible detection of cosmogenic neutrinos will help
 the solution of the UHECR origin when compared to the 
 direct observations.
  
\section{Acknowledgments} 
 The author has benefited from discussions with V.S.~Berezinsky, P.~Blasi,
 D.~DeMarco, T.K.~Gaisser, D.~Seckel, and A.A.~Watson. D.~Allard
 helped me with understanding of the heavy nuclei model of UHECR. 
 This work is supported in part by NASA APT grant NNG04GK86G.  



\end{document}